\documentclass{aastex62}
\usepackage{natbib}
\graphicspath{{./}{figures/}}

\received{xxxx}
\revised{xxxxx}
\accepted{xxxxx}
\submitjournal{ApJL}

\shorttitle{Heavy elements in P-rich stars}
\shortauthors{Masseron et al.}

\begin{document}

\title{Heavy element abundances in P-rich stars: A new site for the $s$-process?}

\correspondingauthor{Thomas Masseron}
\email{tmasseron@iac.es}

\author[0000-0002-6939-0831]{T. Masseron}
\affiliation{Instituto de Astrof\'isica de Canarias, E-38205 La Laguna, Tenerife, Spain}
\affiliation{Departamento de Astrof\'isica, Universidad de La Laguna, E-38206 La Laguna, Tenerife, Spain}

\author[0000-0002-1693-2721]{D. A. García-Hernández}
\affiliation{Instituto de Astrof\'isica de Canarias, E-38205 La Laguna, Tenerife, Spain}
\affiliation{Departamento de Astrof\'isica, Universidad de La Laguna, E-38206 La Laguna, Tenerife, Spain}

\author[0000-0003-2100-1638]{O. Zamora}
\affiliation{Instituto de Astrof\'isica de Canarias, E-38205 La Laguna, Tenerife, Spain}
\affiliation{Departamento de Astrof\'isica, Universidad de La Laguna, E-38206 La Laguna, Tenerife, Spain}

\author[0000-0002-3011-686X]{A. Manchado}
\affiliation{Instituto de Astrof\'isica de Canarias, E-38205 La Laguna, Tenerife, Spain}
\affiliation{Departamento de Astrof\'isica, Universidad de La Laguna, E-38206 La Laguna, Tenerife, Spain}
\affiliation{Consejo Superior de Investigaciones Científicas, Madrid, Spain}

\begin{abstract}
The recently discovered phosphorus-rich stars 
pose a 
challenge to stellar evolution and nucleosynthesis theory, as 
none of the existing models can explain their extremely peculiar chemical abundances pattern. Apart from the 
large phosphorus enhancement, such stars also show enhancement in other light (O, Mg, Si, Al) and heavy (e.g., Ce) 
elements. We have obtained high-resolution optical spectra of two optically bright phosphorus-rich stars 
(including a new P-rich star), for which we have determined a larger number of elemental abundances (from C to Pb). 
We confirm the unusual light-element abundance pattern with very large enhancements of Mg, Si, Al, and P, and 
possibly some Cu enhancement, but the spectra of the new P-rich star is the only one to reveal some C(+N) 
enhancement. When compared to other appropriate metal-poor and neutron-capture enhanced stars, the two P-rich 
stars show heavy-element overabundances similar to low neutron density $s$-process nucleosynthesis, with 
high first- (Sr, Y, Zr) and second-peak (Ba, La, Ce, Nd) element enhancements (even some Pb enhancement 
in one star) and a negative [Rb/Sr] ratio.  However, this $s$-process is distinct from the one occurring 
in asymptotic giant branch (AGB) stars. The notable distinctions encompass larger [Ba/La] and lower Eu and Pb 
than their AGB counterparts. Our observations  should guide stellar nucleosynthesis theoreticians and 
observers to identify the P-rich star progenitor, which represents a new site for $s$-process 
nucleosynthesis, with important implications for the chemical evolution of our Galaxy.

\end{abstract}

\keywords{nuclear reactions, nucleosynthesis, abundances --- stars: abundances --- stars: chemically peculiar --- stars: Population II}

\section{Introduction} \label{sec:intro}
The $s$-process channel for the synthesis of elements beyond Fe (hereafter heavy elements) was formulated as 
early as the pioneering work of \citet{Burbidge1957}. After the observations of $s$-process-rich stars \citep[and 
C-rich; e.g.,][]{Lambert1985,Vanture1992}, it was gradually admitted that this type of 
nucleosynthesis occurs mainly in asymptotic giant branch (AGB) stars \citep[see reviews 
by][]{Busso1999,Herwig2005,Kappeler2011,Karakas2014}.
The principal alternative way to build heavy elements in stars involves $r$-process nucleosynthesis 
with very high neutron densities. In contrast to the $s$-process, the main $r$-process stellar site is still heavily
debated, alternative candidate sites being supernovae or neutron star mergers \citep[e.g.,][]{Argast2004}. In any case, the 
dual nucleosynthetic origin of heavy elements between the $s$- and $r$-processes has found remarkable support
through the observation of very metal-poor stars. Indeed, thanks to their pristine composition, the heavy 
element abundance patterns in these stars clearly separates them into pure $s$-process and pure $r$-process in a 
nearly perfect agreement with theory \citep{Sneden2003,Johnson2004}. Nevertheless, some deviations from 
those standard processes have appeared: first, star-to-star variations of first-peak elements (Sr, Y, Zr) around the 
main $r$-process exist \citep{Truran2002}. This has led theoreticians to invoke the contribution of weak $r$- and 
weak $s$-processes in massive stars \citep{Wasserburg1996,Pignatari2008}. Secondly, a third category of metal-poor 
stars has appeared with an apparently heavy element abundance pattern that is somehow intermediate between the $s$- and 
$r$-processes \citep{Barbuy1997,Jonsell2006}. These observations have led modelers to revive a third  
neutron-capture process with intermediate neutron densities between the $s$- and the $r$-process, the 
so-called $i$-process \citep{Cowan1977}. This process may notably occur in low-metallicity AGB stars 
\citep{Dardelet2014,Hampel2019,Karinkuzhi2020}, but there are other candidate stellar sites (super-AGBs, 
\citealp{Doherty2015}, accreting white dwarfs, \citealp{Denissenkov2017}, and Population~III massive stars, \citealp{Clarkson2018}).

\citet{Masseron2020} have very recently discovered in the near-IR ($H$-band) SDSS-IV/APOGEE-2 survey \citep{Majewski2016} 
a new kind of stars, which show extremely high phosphorus (P) abundances together with high O, Al, Mg, and Si. While 
those authors have discussed and largely rejected almost all possible existing models in order to explain such an 
enhancement of light elements, they have also noted the enhancement of several heavy elements in the only available 
P-rich stellar optical spectrum. \citet{Masseron2020} could not find a satisfying agreement with the theoretical models 
regarding their neutron-capture element yield predictions either. Here, we use a more empirical approach to understanding
the heavy-element nucleosynthesis in P-rich stars. We present a chemical abundance analysis - including a higher 
number of heavy elements (up to Pb) for two P-rich stars - and compare it with those observed in different types of 
metal-poor and neutron capture-rich stars; such comparison stars are chosen to better reflect the variety of neutron-capture processes in stars.

\section{Observational data and chemical analysis}\label{sec:data}
The two sample stars have been primarily identified in the near-IR ($H$-band) by the SDSS-IV/APOGEE-2 survey 
\citep{Majewski2016} with the respective IDs 2M13535604+4437076 and 2M22045404-1148287. While the first star 
is part of the original sample of \citet{Masseron2020}, the second is a newly identified P-rich star. The 
observations and chemical analysis are very similar to those of \citet{Masseron2020} and will not be repeated in detail 
here. In brief, our analysis is based on high-resolution ($R\sim 67\,000$) optical (obtained with the NOT/FIES 
spectrograph) and near-IR $H$-band ($R\sim 22\,500$) spectra (from the APOGEE instrument). The near-IR spectra 
allow the determination of the phosphorus abundances (also S and Co with no useful lines in the optical; 
see Table 1), and therefore the identification of the P-rich stars, but the optical spectra are essential for the 
determination of an extensive number of heavy elements (see Table 1). Unfortunately, the targets stars 
are K giants and are therefore relatively faint in the optical. Consequently, for each star four exposures of one 
hour with the optical NOT/FIES spectrograph were necessary. The NOT/FIES optical observations were obtained 
at three different very recent dates, on 2020 January 12 and 2020 February 20 for 2M13535604+4437076 and on 2020 August 19 for 
2M22045404-1148287. After standard data reduction by the NOT/FIES pipeline and the combination of the four 
exposures, the signal-to-noise (S/N) achieved was 40 and 60 respectively at 5000 \AA. We note, however, 
that we adopted a slightly different strategy regarding the data reduction of 2M22045404-1148287 for the bluest 
part of the spectrum ($\rm<4200\AA$), where the S/N drops drastically, with a range of [3--10] among the several 
exposures. For this particular region (only used for the Pb abundance measurement), we then merged only the 
two best spectra rather than all four of them. This prevented contamination of the spectrum by spurious 
noise artifacts that would have affected our Pb determination. A separate and consistent chemical 
abundance analysis of the two spectra (both optical and near-IR) was then carried out with the BACCHUS code 
\citep{BACCHUS2016}. The effective temperatures, surface gravities, and microturbulence velocities were 
fixed to the calibrated values from the APOGEE DR16 release \citep{Jonsson2020}, while the metallicities and 
abundances were derived by the BACCHUS code. The final atmospheric parameters are $T_{\rm eff}=5143$~K, 
$\log g=2.50$, [Fe/H] $=-1.07$, $\mu_t =1.3$ km/s, and $T_{\rm eff}=4578$~K, $\log g=1.64$, [Fe/H] $=-1.26$, $\mu_t=1.48 $ km/s 
for 2M13535604+4437076  and 2M22045404-1148287 respectively. The full set of derived abundances is presented 
in Table~\ref{tab:abundances}. A lower number of elements was measured in 2M13535604+4437076 because of a 
combination of the lower S/N of the spectra and a higher temperature. We also note that while the analysis of 
the near-IR and optical spectra gives very consistent results concerning the derived abundances, the 
metallicities are offset by $\sim$0.1 dex similarly to what has been observed in globular cluster red giants 
\citep{Masseron2019}, but without a clear explanation to date. In addition, we have compared the individual 
chemical abundance pattern of the two P-rich stars with those of another four metal-poor stars, which are 
representative of the three main neutron-capture nucleosynthesis processes: two CH-stars for the $s$-process 
(HD~26 and HD~206983), one carbon-enhanced metal-poor star with intermediate-process (CEMP-rs or CEMP-i, 
HD~224959), and one extremely metal-poor star with strong $r$-process (CS~22892-052). The stellar parameters 
for the first three stars are extracted from \citet{MasseronPhD}, while we adopt the values from 
\citet{Sneden2003} for the fourth one. The choice of CH-stars as comparison stars for the $s$-process instead 
a CEMP-s star \citep[such as the star of ][]{Johnson2004} was done intentionally as they have a very similar 
metallicity ([Fe/H] $\sim -1.0$) to our two P-rich stars. Regarding the comparison with standard stars of the 
$i$-process, there are no CEMP-rs stars known at the intermediate metallicity of the actually known sample of 
P-rich stars \citep{Masseron2020}. Some post-AGB stars in the Small Magellanic Cloud seem to show such 
$i$-process nucleosynthesis \citep{Hampel2019}, but the number of elements studied is too low to constrain
the following discussion. Therefore, here we assume that the star HD~224959 is a good representative of 
the $i$-process at metallicities [Fe/H] $\sim -1$. As for the $r$-process star representative, it also has a 
much lower metallicity than our P-rich stars. However, it has been demonstrated that the $r$-process in this 
star is identical with the solar $r$-process, thus {\it a priori} making the $r$-process nucleosynthesis 
independent of metallicity. Finally, we stress that there are no high-resolution near-IR spectra available 
for the comparison stars, so we cannot provide their near-IR abundances (e.g., P) in the following plots.

\section{Discussion}\label{sec:discussion}

\begin{figure}[ht!]
\centering
\includegraphics[angle=-90,width=0.8\textwidth]{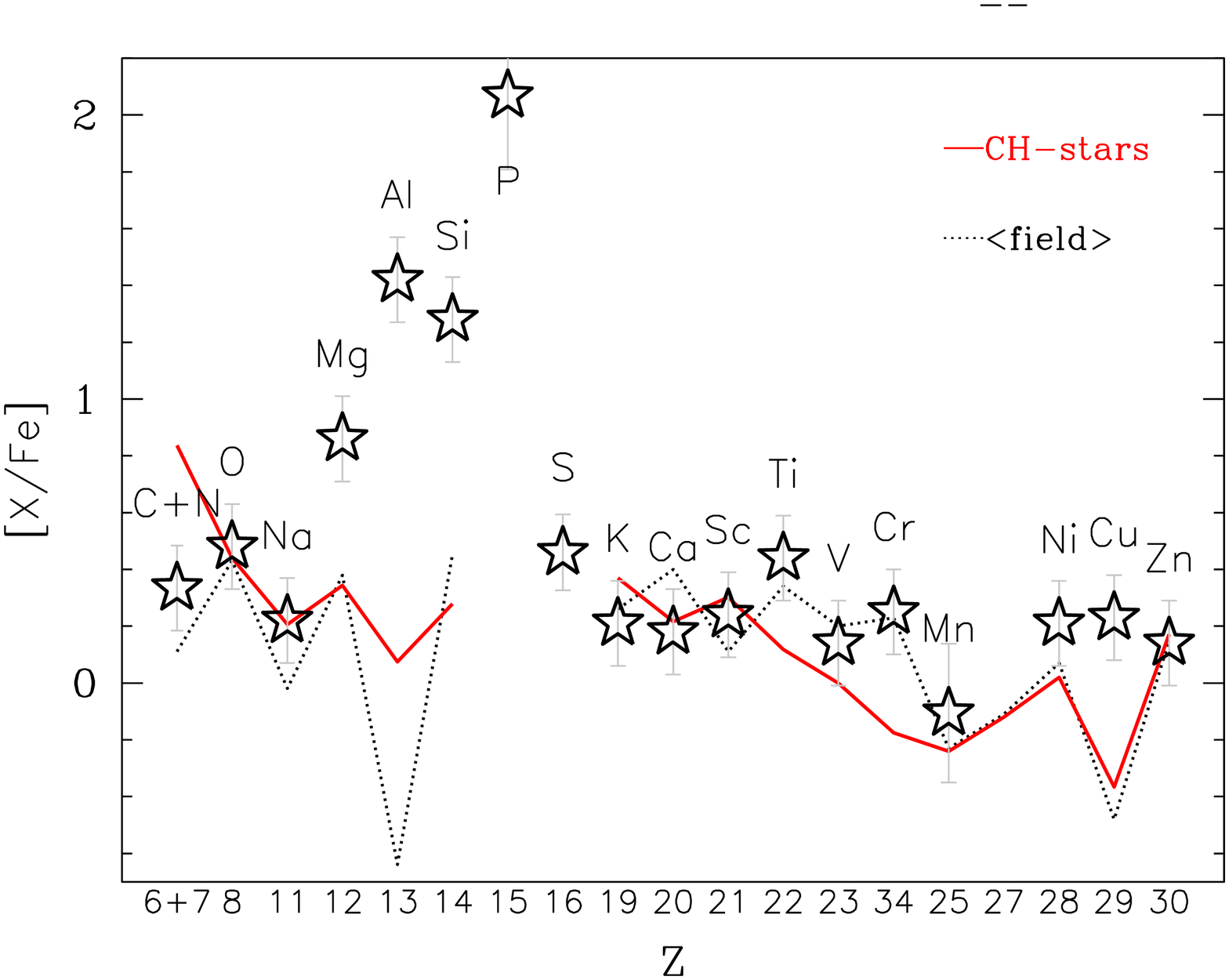}
\includegraphics[angle=-90,width=0.8\textwidth]{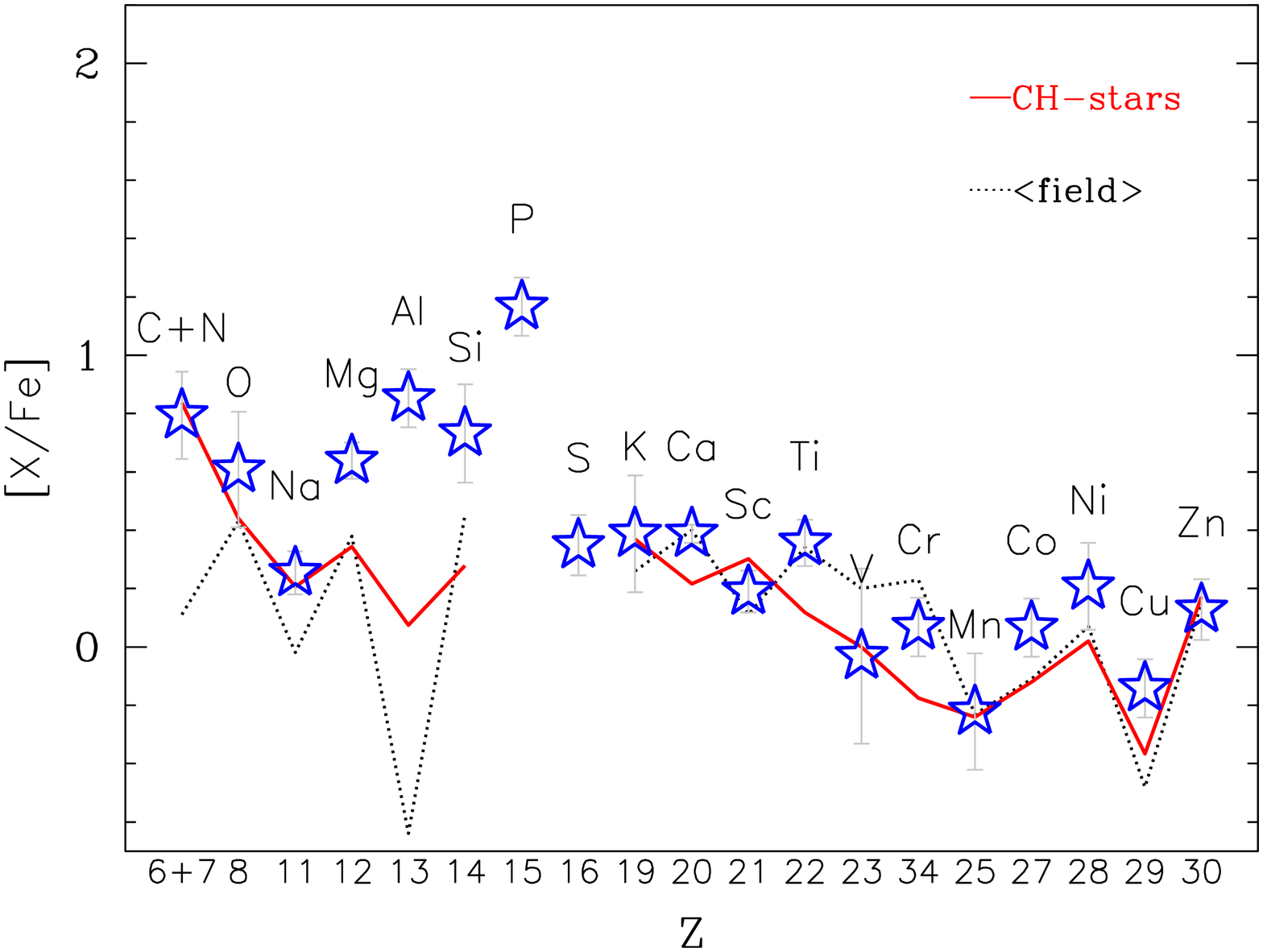}
\caption{Light element abundance pattern for the two P-rich sample stars (upper panel 2M13535604+4437076, 
lower panel 2M22045404-1148287) together with the average abundances of the two CH-stars (red continuous line) 
and the average abundances of field stars with -1.5 $<$ [Fe/H] $<$ -1.0 from \citet{Roederer2014}. Mg, Al, and Si are 
clearly enhanced compared to the CH-stars and field stars. Also, Cu seems to be enhanced in the P-rich stars.}
\label{fig:lightelements}
\end{figure}

In Fig.~\ref{fig:lightelements}, we compare the light element abundances of our two P-rich sample stars with optical 
spectra against the average abundances of the CH-stars and field stars at similar metallicities. First of all, 
the newly observed P-rich star (2M22045404-1148287) shows a chemical abundance pattern very similar to those of the 
other P-rich stars; i.e., it shows a clearly high phosphorus enhancement, as well as strong Mg, Si, and Al compared 
to field stars. Actually, whereas 2M13535604+4437076 shows the largest Mg, Al, P, and Si enhancements among the 
P-rich stars, 2M22045404-1148287 displays more modest enhancements; however, it is not clear whether this is due 
to dilution or to real variations within the nucleosynthesis of the P-rich star progenitors. In contrast, the O 
enhancement is weak in the two P-rich stars (and maybe even not enhanced at all given the error bars) compared 
to the average of the P-rich stars \citep{Masseron2020}. Curiously, we observe a significantly high Cu abundance 
in the two P-rich sample stars; especially in the higher S/N optical spectrum of the P-rich star 2M22045404-1148287. 
The nucleosynthesis of Cu could have various stellar origins \citep{Bisterzo2005,Kobayashi2020}, making very difficult 
the disentanglement of its true origin. Nevertheless, it is very puzzling that no simultaneous Zn enhancement is observed. 

Compared to the CH-star average chemical abundance pattern, it is obvious that Mg, Si, and Al are incompatible, making it 
clear that the progenitors of the CH-stars are not the same as the progenitors of the newly discovered P-rich stars. We 
also remind the reader that none of the P-rich stars shows radial velocity variations over several years of observations 
\citep{Masseron2020}, which is another clear distinction from the CH-stars. However, 2M22045404-1148287 is the first 
P-rich star to show C(+N) enhancement; at a similar level to that of the CH-stars. As for C, it is very difficult to interpret 
the real reason of such an apparent enhancement, as many kinds of stars can produce C, but we may relate this to the fact 
that 2M22045404-1148287 is the lowest metallicity star ([Fe/H] $=-1.23$) among the whole P-rich star sample studied so far 
\citep[15 stars, see][]{Masseron2020}. Therefore, we may tentatively invoke a metallicity effect in the P-rich star 
progenitor nucleosynthesis, which would only allow a significant C enhancement at the lowest metallicities. 

\begin{figure}[ht!]
\centering
\includegraphics[angle=-90,width=0.8\textwidth]{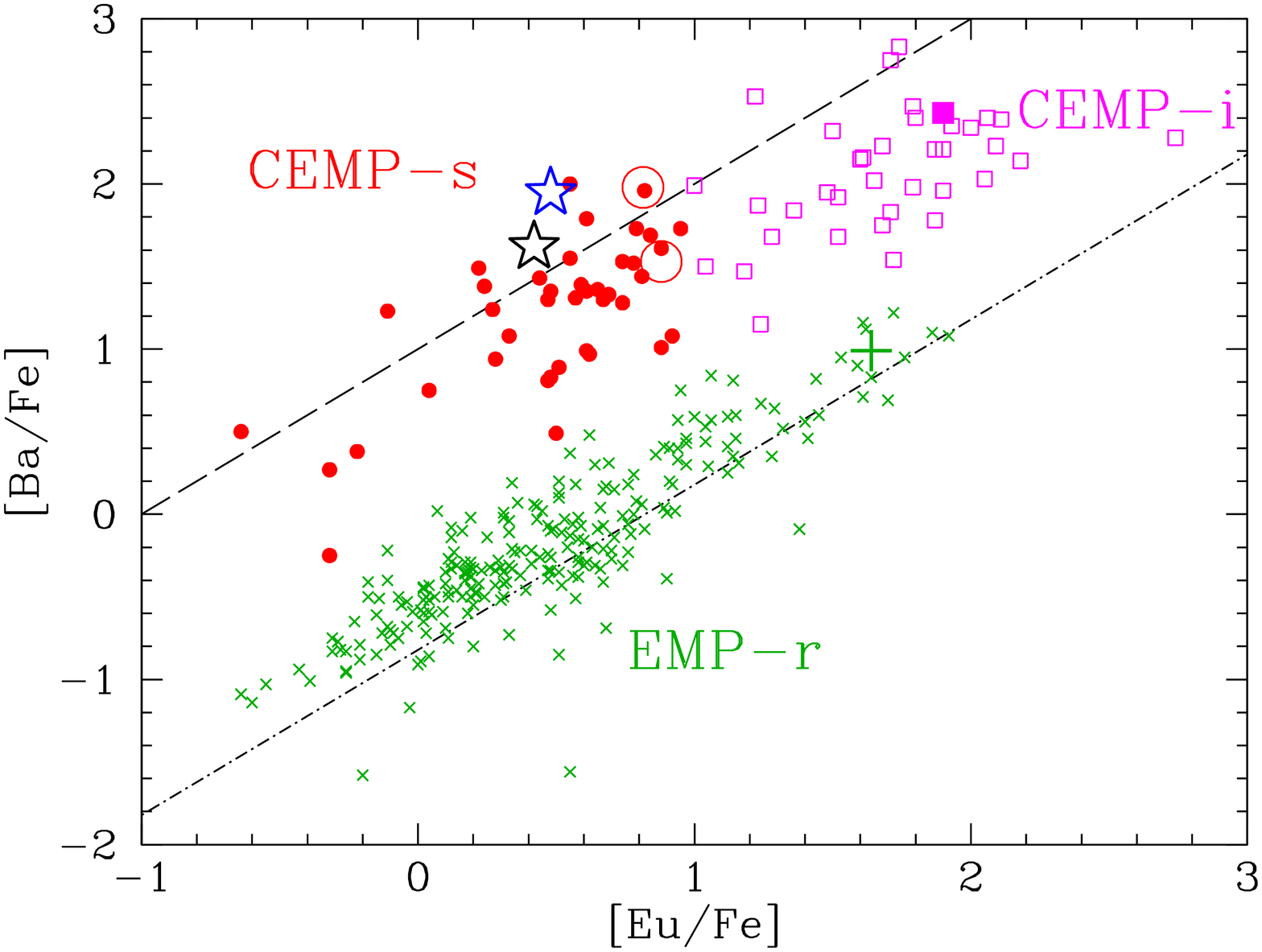}
\caption{Ba versus Eu for very metal-poor stars. The two open starred symbols are the two P-rich stars studied here. 
The two red open circles are the CH-stars HD~26 and HD~206983, the magenta solid square is the CEMP-rs star HD~224959 
and the plus sign the $r$-process-rich star CS~22892-052. All other symbols represent very metal-poor stars ([Fe/H] $< -2.0$) 
from \citet{SAGA2008}. The definition for the various metal-poor stars sub-categories (CEMP-i, CEMP-s, EMP-r) has been 
adopted from \citet{Masseron2010}. The dashed line corresponds to the theoretical pure $s$-process prediction and the 
dashed-dotted line to the theoretical pure $r$-process prediction. The P-rich star Ba and Eu abundances agree quite 
well with the CEMP-s stars. }\label{fig:BaFevsEuFe}
\end{figure}

Regarding the heavy element nucleosynthesis, we have displayed all the stars (P-rich and comparison stars) in the 
[Ba/Fe] versus [Eu/Fe] diagram (Fig.~\ref{fig:BaFevsEuFe}), as this plane has been demonstrated to be an excellent 
diagnostic for separating the three main neutron-capture processes in metal-poor stars \citep{Masseron2010}. In this 
figure, the P-rich stars appear clearly among the CEMP-s stars, quite close to the CH-stars and the theoretical $s$-process predictions.

The $s$-process signature is confirmed when examining the full distribution of heavy elements (Fig. 3). In this 
figure, we compare the heavy element abundance pattern of the P-rich stars against those of the CH-stars, the CEMP-i 
(or CEMP-rs) star and the $r$-process-rich EMP-r star. It is clear that the P-rich star heavy element pattern is clearly 
incompatible with the $i$- and $r$-processes, particularly when comparing the Eu and Sm abundances. Conversely, 
the heavy element chemical pattern of the P-rich stars follows that of the CH-stars and is remarkably close 
in the case of 2M22045404-1148287. This strongly suggests that the P-rich star progenitors have undergone some
 kind of $s$-process nucleosynthesis. We note that for an unbiased comparison of the s-process it is particularly 
important that the P-rich stars and the CH-stars have similar metallicities. Indeed, metallicity is known to play 
a key-role in $s$-process nucleosynthesis as it affects the second-to-first peak element ratio \citep{Gallino1998, 
Goriely2000}. The second-to-first peak element ratio matches remarkably well the CH-stars and the P-rich 
star 2M22045404-1148287 but it is less obvious with the slightly higher metallicity P-rich star 2M13535604+4437076. 
This discrepancy between the two P-rich stars could be related to a difference in the $s$-process strength \citep{Gallino1998, Goriely2000}.

There are some more element-specific differences between the CH-stars and the P-rich stars that can further constrain 
the details of the $s$-process nucleosynthesis that has occurred in the P-rich star progenitors. As already 
noticed by \citet{Masseron2020}, the Ba overabundance is particularly large compared to the other second-peak elements 
(e.g., La) and the corresponding CH-star values. Our differential approach with the CH-stars, displaying similar 
stellar parameters, confirms that the large Ba enhancement (and hence the high [Ba/La] ratio) is real and not due to NLTE 
or 3D effects. Such a large discrepancy between the two contiguous elements Ba and La is unusual and may provide 
key clues regarding the details (e.g., neutron density and exposure) of the nucleosynthesis process; something that 
future detailed nuclear network simulations could clarify. Nevertheless, the [Rb/Sr] ratio being negative in
 both cases also puts quite a low upper limit to the neutron density (i.e., $\lesssim 10^{11}$cm$\rm ^{-3}$). Moreover, 
although both P-rich stars have negative [Rb/Sr], the value for the star 2M22045404-1148287 is significantly smaller 
than that for 2M13535604+4437076. Taking into account that the [Rb/Sr] (or [Rb/Zr]) ratio is sensitive to the neutron 
density \citep[see e.g.,][and references therein]{garcia-hdez2006}, this implies that the neutron density is larger in 
2M13535604+4437076.

When looking at the third-peak elements, one P-rich sample star (2M22045404-1148287, the only star where we could 
detect the Pb lines in the bluest optical region; see Fig.~\ref{fig:Pblines}) shows some Pb enhancement \citep[consistently, 
in principle, with the predictions for the $s$-process in AGB stars;][]{VanEck2003}. Interestingly, the Pb abundance in the 
P-rich star is lower than in the CH-stars. This relatively modest Pb enrichment may be another indication that the P-rich 
star progenitors have undergone an $s$-process nucleosynthesis with distinct characteristics. 

The $s$-process is believed to occur only in AGB stars and there are no other stellar evolution and nucleosynthesis models 
in the literature where the $s$-process occurs. Fortunately, variations in the details of the $s$-process are still allowed by theory. 
In their nuclear network simulations at similar metallicities to those of our P-rich stars, \citet{Hampel2019} explore the [Pb/Ba] 
and [Ba/Sr] ratios with varying neutron density and exposure. By assuming that the temperatures and gas densities in the 
region where neutron-capture elements are formed in the P-rich star progenitors are similar for AGB stars, we deduce - 
from their work (see e.g. their Fig. 1) and our Rb, Sr, Ba, and Pb abundances - that for a neutron density below 10$\rm^{11}$, 
the neutron exposure in the P-rich stars is between 0.7 and 1.2 mbarn$\rm^{-1}$.

Moreover, while the $s$-process is also capable of making Cu, it is expected to make Zn as well. But neither of the two P-rich 
stars shows any Zn enhancement, implying that Cu has not been enhanced by such a neutron-capture process. 

Finally, we remind the reader that $^{31}$P can also be produced by the $s$-process, thanks to the large neutron cross section of 
$^{30}$Si. In AGB and super-AGB star models at [Fe/H] $\sim -1$ \citep{Karakas2010, Doherty2015}, phosphorus production is 
negligible. Although the light-element chemical pattern in P-rich stars has already ruled out an AGB scenario \citep{Masseron2020}, 
it could be still possible that P may have a neutron-capture origin in the P-rich star progenitors, especially given that Si is
 strongly enhanced as well. Therefore, more extensive nuclear network simulations of neutron-capture processes with various 
neutron exposures and densities are strongly encouraged in order to explore the tight observational constraints reported here, 
with special attention to the [Rb/Sr], [Ba/La], [La/Eu], and [Pb/La] ratios, along with P production and possibly the [Cu/Zn] ratio.


\begin{figure}[ht!]
\centering
\includegraphics[angle=-90,width=0.8\textwidth]{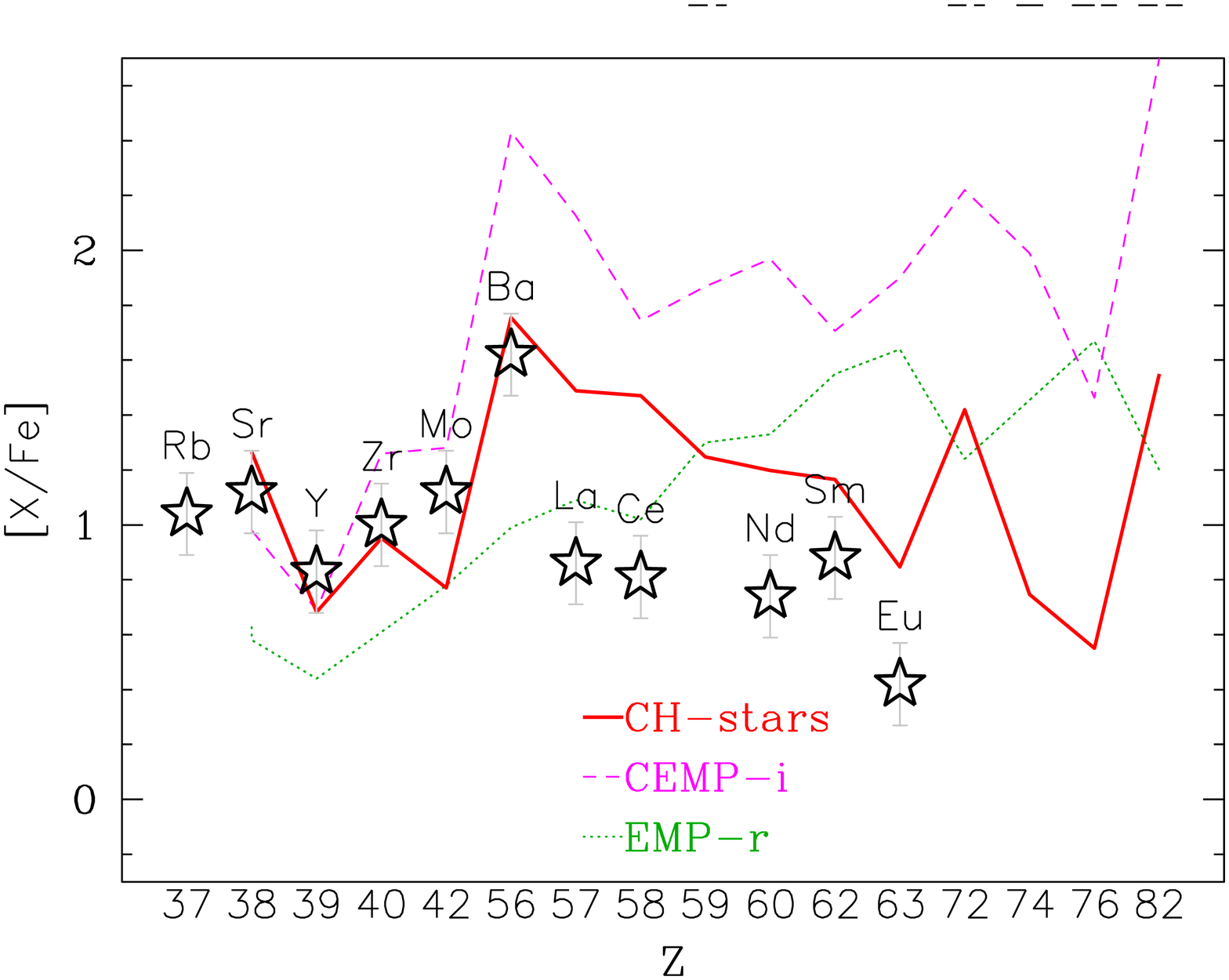}
\includegraphics[angle=-90,width=0.8\textwidth]{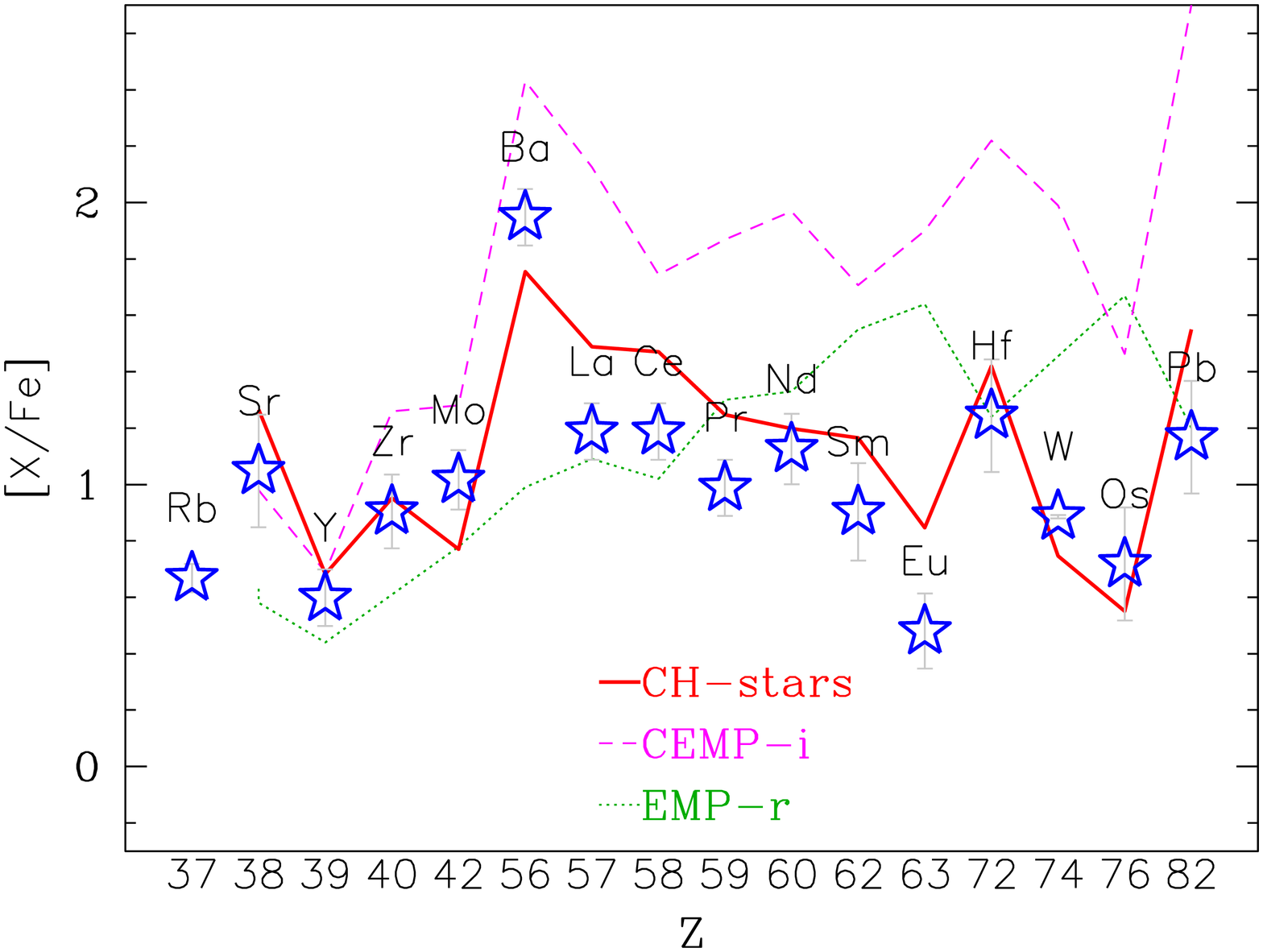}
\caption{Heavy element abundance pattern for the two P-rich sample stars (upper panel 2M13535604+4437076, lower panel 
2M22045404-1148287), together with the average abundance of the two CH-stars (red continuous line), the abundance pattern 
of the CEMP-i star HD~224959 (magenta dashed line), and the abundance pattern of the r-process-rich star CS22892-052 (dotted 
green line). The heavy element abundance pattern observed in P-rich stars is more similar to the chemical pattern of the CH-stars.}
\label{fig:heavyelements}
\end{figure}

\begin{figure}[ht!]
\centering
\includegraphics[angle=-90,width=0.8\textwidth]{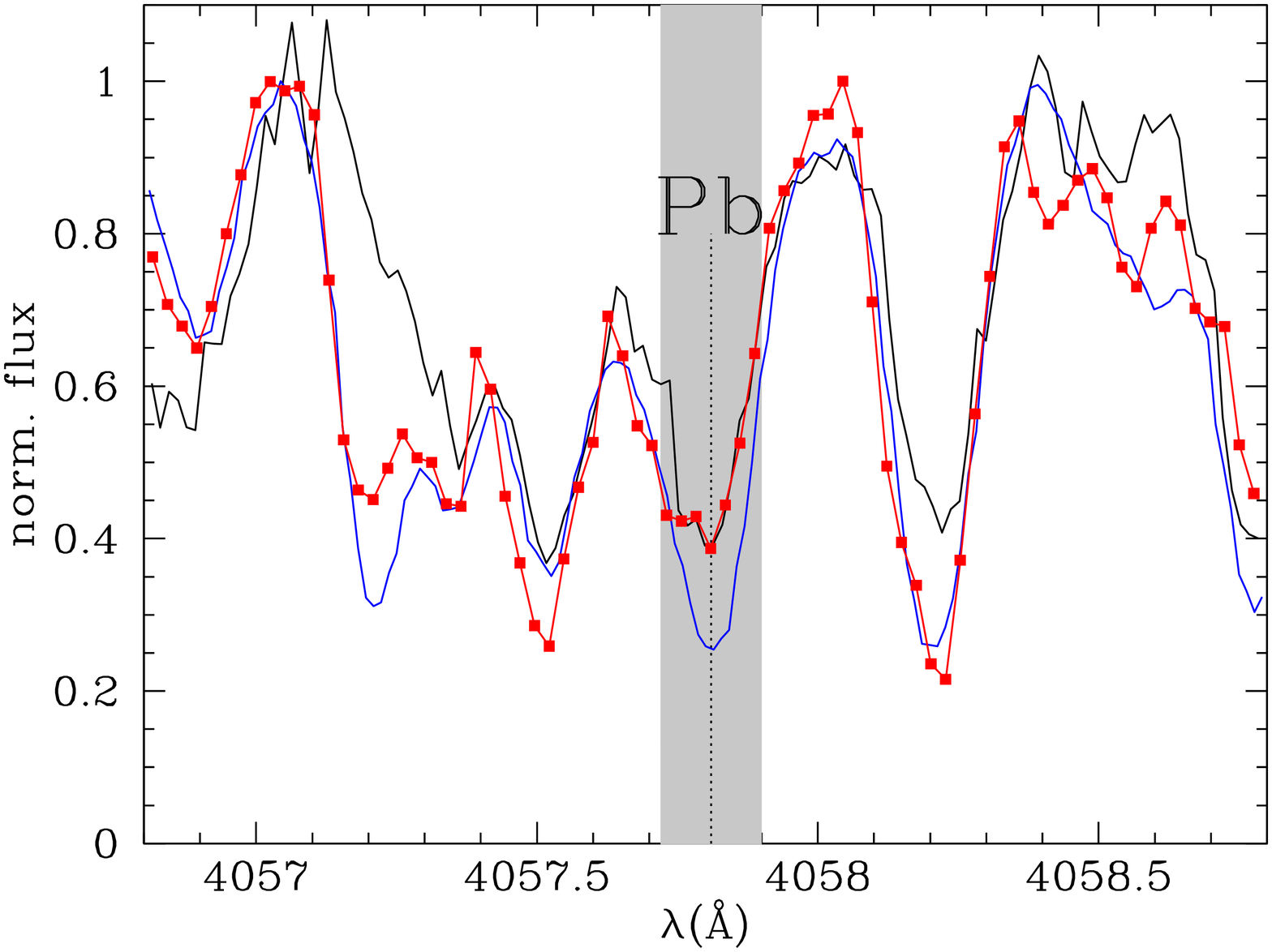}
\includegraphics[angle=-90,width=0.8\textwidth]{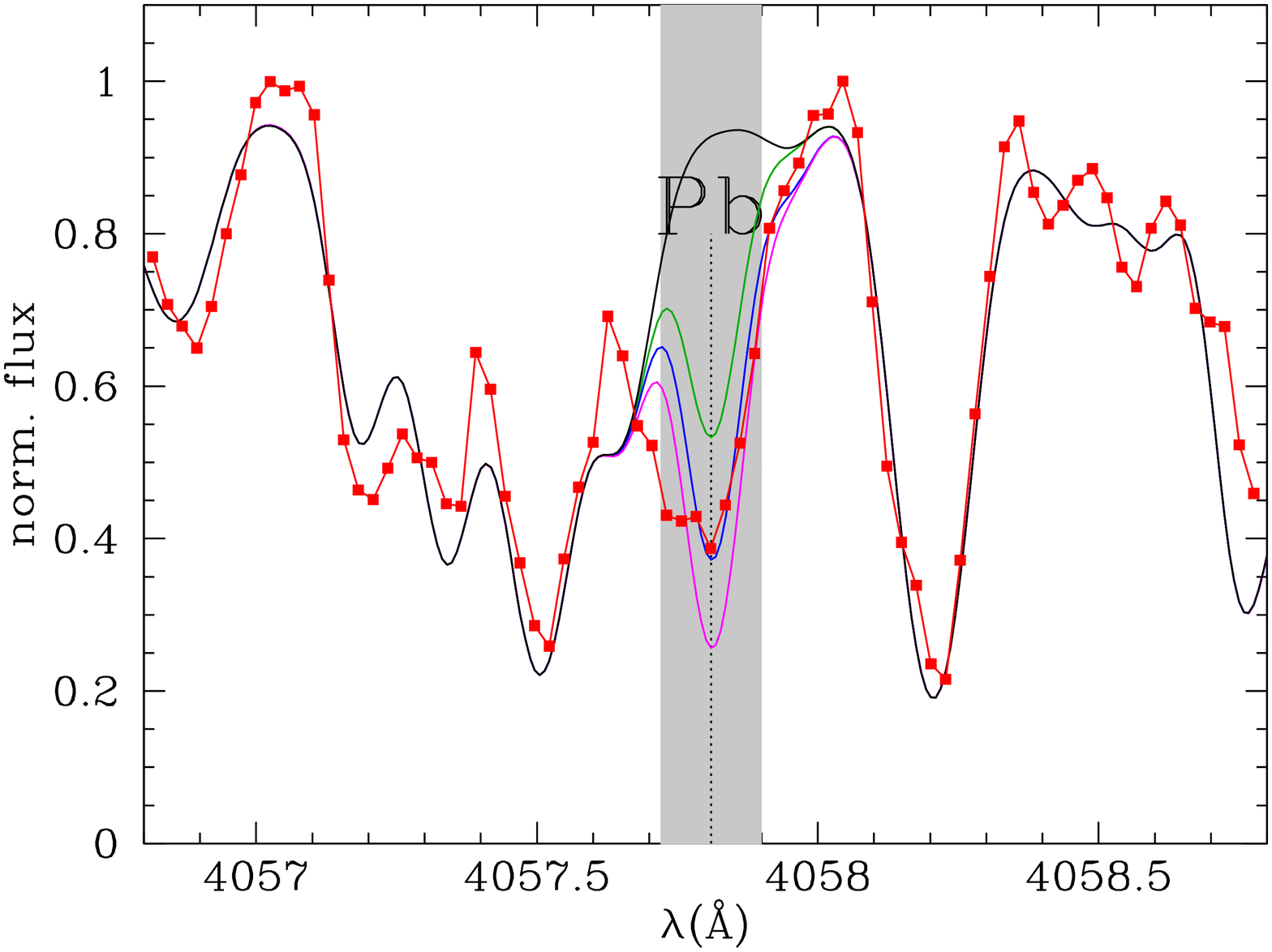}
\caption{Pb line at 4057.8~\AA~in the P-rich star 2M22045404-1148287 optical spectrum. Upper panel: the red line with dots 
corresponds to the observed  P-rich star spectrum, and blue and black continuous line correspond to, respectively, HD~206983 
and HD~26 spectra. Lower panel: the red line with dots corresponds to the P-rich star spectrum and the continuous lines 
correspond to the synthesis without Pb (black), with our best Pb determination (blue), with our best Pb abundance 
minus 0.5 dex (green) and with the same abundances as in the CH-stars (magenta). The Pb line is blended by a CH line on the 
blue side. The optical spectrum of the P-rich star in the bluest region displays a rather low S/N ($\sim$15), but the Pb 
line is clearly detected.  }\label{fig:Pblines}
\end{figure}

\section{Concluding remarks}

Our detailed study of the heavy element abundance pattern of two optically bright P-rich stars lead us to conclude 
that it is very likely that the P-rich stars progenitors have undergone a kind of $s$-process nucleosynthesis with 
potentially a low neutron density and low neutron exposure. However, their light element abundances do not 
correspond to AGB stars; the only stellar site currently known where such neutron-capture nucleosynthesis takes place. 
Therefore, whenever a new stellar site or channel for the $s$-process is identified, it will provide clues on the
 nature of the P-rich star progenitors and their role in the context of Galactic chemical evolution. Thus, our observations 
should guide the stellar nucleosynthesis modelers. On the observational side, our slightly enhanced Pb and (probably) 
Cu abundances in P-rich stars, should be completely confirmed with more  high-resolution optical 
spectroscopic P-rich star observations and/or much better S/N spectra.

\acknowledgements
The authors acknowledge support from the State Research Agency (AEI) of the Spanish Ministry of Science, Innovation 
and Universities (MCIU) and the European Regional Development Fund (FEDER) under grant AYA2017-88254-P. This article 
is based on observations made at the IAC'S Observatorios de Canarias with the Nordic Optical Telescope (NOT) operated 
on the island of La Palma by NOTSA at Roque de los Muchachos Observatory (ORM). Funding for the Sloan Digital Sky Survey 
IV has been provided by the Alfred P. Sloan Foundation, the U.S. Department of Energy Office of Science, and the 
Participating Institutions. SDSS acknowledges support and resources from the Center for High-Performance Computing 
at the University of Utah. SDSS is managed by the Astrophysical Research Consortium for the Participating Institutions 
of the SDSS Collaboration including the Brazilian Participation Group, the Carnegie Institution for Science, Carnegie 
Mellon University, the Chilean Participation Group, the French Participation Group, Harvard-Smithsonian Center for 
Astrophysics, Instituto de Astrofísica de Canarias, The Johns Hopkins University, Kavli Institute for the Physics 
and Mathematics of the Universe (IPMU) / University of Tokyo, the Korean Participation Group, Lawrence Berkeley National 
Laboratory, Leibniz Institut für Astrophysik Potsdam (AIP), Max-Planck-Institut für Astronomie (MPIA Heidelberg), 
Max-Planck-Institut für Astrophysik (MPA Garching), Max-Planck-Institut für Extraterrestrische Physik (MPE), National
 Astronomical Observatories of China, New Mexico State University, New York University, University of Notre Dame, 
Observatório Nacional/MCTI, The Ohio State University, Pennsylvania State University, Shanghai Astronomical Observatory, 
United Kingdom Participation Group, Universidad Nacional Autónoma de México, University of Arizona, University of
 Colorado Boulder, University of Oxford, University of Portsmouth, University of Utah, University of Virginia, 
University of Washington, University of  Wisconsin, Vanderbilt University, and Yale University.
 
\bibliographystyle{aasjournal}
\bibliography{Prichheavyelements}

\begin{thebibliography}{}
\expandafter\ifx\csname natexlab\endcsname\relax\def\natexlab#1{#1}\fi
\providecommand{\url}[1]{\href{#1}{#1}}

\bibitem[{{Argast} {et~al.}(2004){Argast}, {Samland}, {Thielemann}, \&
  {Qian}}]{Argast2004}
{Argast}, D., {Samland}, M., {Thielemann}, F.~K., \& {Qian}, Y.~Z. 2004, \aap,
  416, 997

\bibitem[{{Asplund} {et~al.}(2009){Asplund}, {Grevesse}, {Sauval}, \&
  {Scott}}]{Asplund2009}
{Asplund}, M., {Grevesse}, N., {Sauval}, A.~J., \& {Scott}, P. 2009, \araa, 47,
  481

\bibitem[{{Barbuy} {et~al.}(1997){Barbuy}, {Cayrel}, {Spite}, {Beers}, {Spite},
  {Nordstroem}, \& {Nissen}}]{Barbuy1997}
{Barbuy}, B., {Cayrel}, R., {Spite}, M., {et~al.} 1997, \aap, 317, L63

\bibitem[{{Bisterzo} {et~al.}(2005){Bisterzo}, {Pompeia}, {Gallino},
  {Pignatari}, {Cunha}, {Heger}, \& {Smith}}]{Bisterzo2005}
{Bisterzo}, S., {Pompeia}, L., {Gallino}, R., {et~al.} 2005, \nphysa, 758, 284

\bibitem[{{Burbidge} {et~al.}(1957){Burbidge}, {Burbidge}, {Fowler}, \&
  {Hoyle}}]{Burbidge1957}
{Burbidge}, E.~M., {Burbidge}, G.~R., {Fowler}, W.~A., \& {Hoyle}, F. 1957,
  Reviews of Modern Physics, 29, 547

\bibitem[{{Busso} {et~al.}(1999){Busso}, {Gallino}, \&
  {Wasserburg}}]{Busso1999}
{Busso}, M., {Gallino}, R., \& {Wasserburg}, G.~J. 1999, \araa, 37, 239

\bibitem[{{Clarkson} {et~al.}(2018){Clarkson}, {Herwig}, \&
  {Pignatari}}]{Clarkson2018}
{Clarkson}, O., {Herwig}, F., \& {Pignatari}, M. 2018, \mnras, 474, L37

\bibitem[{{Cowan} \& {Rose}(1977)}]{Cowan1977}
{Cowan}, J.~J., \& {Rose}, W.~K. 1977, \apj, 212, 149

\bibitem[{{Dardelet} {et~al.}(2014){Dardelet}, {Ritter}, {Prado}, {Heringer},
  {Higgs}, {Sandalski}, {Jones}, {Denisenkov}, {Venn}, {Bertolli}, {Pignatari},
  {Woodward}, \& {Herwig}}]{Dardelet2014}
{Dardelet}, L., {Ritter}, C., {Prado}, P., {et~al.} 2014, in XIII Nuclei in the
  Cosmos (NIC XIII), 145

\bibitem[{{Denissenkov} {et~al.}(2017){Denissenkov}, {Herwig}, {Battino},
  {Ritter}, {Pignatari}, {Jones}, \& {Paxton}}]{Denissenkov2017}
{Denissenkov}, P.~A., {Herwig}, F., {Battino}, U., {et~al.} 2017, \apjl, 834,
  L10

\bibitem[{{Doherty} {et~al.}(2015){Doherty}, {Gil-Pons}, {Siess}, {Lattanzio},
  \& {Lau}}]{Doherty2015}
{Doherty}, C.~L., {Gil-Pons}, P., {Siess}, L., {Lattanzio}, J.~C., \& {Lau}, H.
  H.~B. 2015, \mnras, 446, 2599

\bibitem[{{Gallino} {et~al.}(1998){Gallino}, {Arlandini}, {Busso}, {Lugaro},
  {Travaglio}, {Straniero}, {Chieffi}, \& {Limongi}}]{Gallino1998}
{Gallino}, R., {Arlandini}, C., {Busso}, M., {et~al.} 1998, \apj, 497, 388

\bibitem[{{Garc{\'\i}a-Hern{\'a}ndez}
  {et~al.}(2006){Garc{\'\i}a-Hern{\'a}ndez}, {Garc{\'\i}a-Lario}, {Plez},
  {D'Antona}, {Manchado}, \& {Trigo-Rodr{\'\i}guez}}]{garcia-hdez2006}
{Garc{\'\i}a-Hern{\'a}ndez}, D.~A., {Garc{\'\i}a-Lario}, P., {Plez}, B.,
  {et~al.} 2006, Science, 314, 1751

\bibitem[{{Goriely} \& {Mowlavi}(2000)}]{Goriely2000}
{Goriely}, S., \& {Mowlavi}, N. 2000, \aap, 362, 599

\bibitem[{{Hampel} {et~al.}(2019){Hampel}, {Karakas}, {Stancliffe}, {Meyer}, \&
  {Lugaro}}]{Hampel2019}
{Hampel}, M., {Karakas}, A.~I., {Stancliffe}, R.~J., {Meyer}, B.~S., \&
  {Lugaro}, M. 2019, \apj, 887, 11

\bibitem[{{Herwig}(2005)}]{Herwig2005}
{Herwig}, F. 2005, \araa, 43, 435

\bibitem[{{Johnson} \& {Bolte}(2004)}]{Johnson2004}
{Johnson}, J.~A., \& {Bolte}, M. 2004, \apj, 605, 462

\bibitem[{{Jonsell} {et~al.}(2006){Jonsell}, {Barklem}, {Gustafsson},
  {Christlieb}, {Hill}, {Beers}, \& {Holmberg}}]{Jonsell2006}
{Jonsell}, K., {Barklem}, P.~S., {Gustafsson}, B., {et~al.} 2006, \aap, 451,
  651

\bibitem[{{J{\"o}nsson} {et~al.}(2020){J{\"o}nsson}, {Holtzman}, {Allende
  Prieto}, {Cunha}, {Garc{\'\i}a-Hern{\'a}ndez}, {Hasselquist}, {Masseron},
  {Osorio}, {Shetrone}, {Smith}, {Stringfellow}, {Bizyaev}, {Edvardsson},
  {Majewski}, {M{\'e}sz{\'a}ros}, {Souto}, {Zamora}, {Beaton}, {Bovy}, {Donor},
  {Pinsonneault}, {Poovelil}, \& {Sobeck}}]{Jonsson2020}
{J{\"o}nsson}, H., {Holtzman}, J.~A., {Allende Prieto}, C., {et~al.} 2020, \aj,
  160, 120

\bibitem[{{K{\"a}ppeler} {et~al.}(2011){K{\"a}ppeler}, {Gallino}, {Bisterzo},
  \& {Aoki}}]{Kappeler2011}
{K{\"a}ppeler}, F., {Gallino}, R., {Bisterzo}, S., \& {Aoki}, W. 2011, Reviews
  of Modern Physics, 83, 157

\bibitem[{{Karakas}(2010)}]{Karakas2010}
{Karakas}, A.~I. 2010, \mnras, 403, 1413

\bibitem[{{Karakas} \& {Lattanzio}(2014)}]{Karakas2014}
{Karakas}, A.~I., \& {Lattanzio}, J.~C. 2014, \pasa, 31, e030

\bibitem[{{Karinkuzhi} {et~al.}(2020){Karinkuzhi}, {Van Eck}, {Goriely},
  {Siess}, \& {Jorissen}}]{Karinkuzhi2020}
{Karinkuzhi}, D., {Van Eck}, S., {Goriely}, S., {Siess}, L., \& {Jorissen}, A.
  2020, \aap, submitted

\bibitem[{{Kobayashi} {et~al.}(2020){Kobayashi}, {Karakas}, \&
  {Lugaro}}]{Kobayashi2020}
{Kobayashi}, C., {Karakas}, A.~I., \& {Lugaro}, M. 2020, \apj, 900, 179

\bibitem[{{Lambert}(1985)}]{Lambert1985}
{Lambert}, D.~L. 1985, {The chemical composition of cool stars: I - The barium
  stars. (Review paper)}, ed. M.~{Jaschek} \& P.~C. {Keenan}, Vol. 114, 191

\bibitem[{{Majewski} {et~al.}(2016){Majewski}, {APOGEE Team}, \& {APOGEE-2
  Team}}]{Majewski2016}
{Majewski}, S.~R., {APOGEE Team}, \& {APOGEE-2 Team}. 2016, Astronomische
  Nachrichten, 337, 863

\bibitem[{{Masseron}(2006)}]{MasseronPhD}
{Masseron}, T. 2006, PhD thesis, Observatoire de Paris, France

\bibitem[{{Masseron} {et~al.}(2020){Masseron}, {Garc{\'\i}a-Hern{\'a}ndez},
  {Santove{\~n}a}, {Manchado}, {Zamora}, {Manteiga}, \&
  {Dafonte}}]{Masseron2020}
{Masseron}, T., {Garc{\'\i}a-Hern{\'a}ndez}, D.~A., {Santove{\~n}a}, R.,
  {et~al.} 2020, Nature Communications, 11, 3759

\bibitem[{{Masseron} {et~al.}(2010){Masseron}, {Johnson}, {Plez}, {van Eck},
  {Primas}, {Goriely}, \& {Jorissen}}]{Masseron2010}
{Masseron}, T., {Johnson}, J.~A., {Plez}, B., {et~al.} 2010, \aap, 509, A93

\bibitem[{{Masseron} {et~al.}(2016){Masseron}, {Merle}, \&
  {Hawkins}}]{BACCHUS2016}
{Masseron}, T., {Merle}, T., \& {Hawkins}, K. 2016, {BACCHUS: Brussels
  Automatic Code for Characterizing High accUracy Spectra}, Astrophysics Source
  Code Library, , , ascl:1605.004

\bibitem[{{Masseron} {et~al.}(2019){Masseron}, {Garc{\'\i}a-Hern{\'a}ndez},
  {M{\'e}sz{\'a}ros}, {Zamora}, {Dell'Agli}, {Allende Prieto}, {Edvardsson},
  {Shetrone}, {Plez}, {Fern{\'a}ndez-Trincado}, {Cunha}, {J{\"o}nsson},
  {Geisler}, {Beers}, \& {Cohen}}]{Masseron2019}
{Masseron}, T., {Garc{\'\i}a-Hern{\'a}ndez}, D.~A., {M{\'e}sz{\'a}ros}, S.,
  {et~al.} 2019, \aap, 622, A191

\bibitem[{{Pignatari} {et~al.}(2008){Pignatari}, {Gallino}, {Meynet},
  {Hirschi}, {Herwig}, \& {Wiescher}}]{Pignatari2008}
{Pignatari}, M., {Gallino}, R., {Meynet}, G., {et~al.} 2008, \apjl, 687, L95

\bibitem[{{Roederer} {et~al.}(2014){Roederer}, {Preston}, {Thompson},
  {Shectman}, {Sneden}, {Burley}, \& {Kelson}}]{Roederer2014}
{Roederer}, I.~U., {Preston}, G.~W., {Thompson}, I.~B., {et~al.} 2014, \aj,
  147, 136

\bibitem[{{Sneden} {et~al.}(2003){Sneden}, {Cowan}, {Lawler}, {Ivans},
  {Burles}, {Beers}, {Primas}, {Hill}, {Truran}, {Fuller}, {Pfeiffer}, \&
  {Kratz}}]{Sneden2003}
{Sneden}, C., {Cowan}, J.~J., {Lawler}, J.~E., {et~al.} 2003, \apj, 591, 936

\bibitem[{{Suda} {et~al.}(2008){Suda}, {Katsuta}, {Yamada}, {Suwa}, {Ishizuka},
  {Komiya}, {Sorai}, {Aikawa}, \& {Fujimoto}}]{SAGA2008}
{Suda}, T., {Katsuta}, Y., {Yamada}, S., {et~al.} 2008, \pasj, 60, 1159

\bibitem[{{Truran} {et~al.}(2002){Truran}, {Cowan}, {Pilachowski}, \&
  {Sneden}}]{Truran2002}
{Truran}, J.~W., {Cowan}, J.~J., {Pilachowski}, C.~A., \& {Sneden}, C. 2002,
  \pasp, 114, 1293

\bibitem[{{Van Eck} {et~al.}(2003){Van Eck}, {Goriely}, {Jorissen}, \&
  {Plez}}]{VanEck2003}
{Van Eck}, S., {Goriely}, S., {Jorissen}, A., \& {Plez}, B. 2003, \aap, 404,
  291

\bibitem[{{Vanture}(1992)}]{Vanture1992}
{Vanture}, A.~D. 1992, \aj, 104, 1997

\bibitem[{{Wasserburg} {et~al.}(1996){Wasserburg}, {Busso}, \&
  {Gallino}}]{Wasserburg1996}
{Wasserburg}, G.~J., {Busso}, M., \& {Gallino}, R. 1996, \apjl, 466, L109

\end{thebibliography}

\begin{table}[ht]
\begin{center}
\begin{tabular}{l|cc|cc|c}

   &  \multicolumn{2}{c|}{2M13535604+4437076} & \multicolumn{2}{c|}{2M22045404-1148287} & inst.\\
  & [X/Fe] & $\sigma$  &  [X/Fe] & $\sigma$ & \\
    \hline
 C &  0.83 &  0.16 &   0.35 &  0.16 &FIES\\ 
 N &  0.57 &  0.04 &   0.26 &  0.04 &FIES\\ 
C+N &  0.79 &  0.15 &   0.33 &  0.15 &FIES\\ 
 O &  0.61 &  0.20 &   0.48 &  0.20 &FIES\\ 
Na &  0.25 &  0.07 &   0.22 &  0.07 &FIES\\ 
Mg &  0.64 &  0.06 &   0.86 &  0.06 &FIES\\ 
Al &  0.85 &  0.10 &   1.42 &  0.10 &FIES\\ 
Si &  0.73 &  0.17 &   1.28 &  0.17 &FIES\\ 
 P &  1.17 &  0.10 &   2.07 &  0.26 &APOGEE\\ 
 S &  0.35 &  0.10 &   0.46 &  0.13 &APOGEE\\ 
 K &  0.39 &  0.20 &   0.21 &  0.20 &FIES\\ 
Ca &  0.39 &  0.03 &   0.18 &  0.03 &FIES\\ 
Sc &  0.19 &  0.07 &   0.24 &  0.07 &FIES\\ 
Ti &  0.36 &  0.08 &   0.44 &  0.08 &FIES\\ 
 V & -0.03 &  0.30 &   0.14 &  0.30 &FIES\\ 
Cr &  0.07 &  0.10 &   0.25 &  0.10 &FIES\\ 
Mn & -0.22 &  0.20 &  -0.11 &  0.24 &FIES\\ 
Co &  0.07 &  0.10 &   $\ldots$ &  $\ldots$ &APOGEE\\ 
Ni &  0.21 &  0.15 &   0.21 &  0.15 &FIES\\ 
Cu & -0.14 &  0.10 &   0.23 &  0.10 &FIES\\ 
Zn &  0.13 &  0.10 &   0.14 &  0.10 &FIES\\ 
Rb &  0.67 &  0.05 &   1.04 &  0.05 &FIES\\ 
Sr &  1.05 &  0.20 &   1.12 &  0.20 &FIES\\ 
 Y &  0.60 &  0.10 &   0.83 &  0.10 &FIES\\ 
Zr &  0.90 &  0.13 &   1.00 &  0.13 &FIES\\ 
Mo &  1.02 &  0.11 &   1.12 &  0.11 &FIES\\ 
Ba &  1.95 &  0.10 &   1.62 &  0.10 &FIES\\ 
La &  1.19 &  0.10 &   0.86 &  0.10 &FIES\\ 
Ce &  1.19 &  0.10 &   0.81 &  0.10 &FIES\\ 
Pr &  0.99 &  0.10 &   $\ldots$ &  $\ldots$ &FIES\\
Nd &  1.13 &  0.13 &   0.74 &  0.13 &FIES\\ 
Sm &  0.90 &  0.17 &   0.88 &  0.17 &FIES\\ 
Eu &  0.48 &  0.13 &   0.42 &  0.13 &FIES\\ 
Hf &  1.24 &  0.20 &   $\ldots$ &  $\ldots$ &FIES\\ 
 W &  0.89 &  0.01 &   $\ldots$ &  $\ldots$ &FIES\\ 
Os &  0.72 &  0.20 &   $\ldots$ &  $\ldots$ &FIES\\ 
Pb &  $\ldots$ &  $\ldots$ &   1.17 &  0.20 &FIES\\ 

    \hline
\end{tabular}
\end{center}
\caption{Elemental abundances for the two P-rich stars and their respective errors computed with the 
line-to-line rms or the fit residual when only one line is used. The solar reference abundance is 
\citet{Asplund2009}. The last column indicates from which spectrograph the abundance has been derived.}
\label{tab:abundances}
\end{table}

\end{document}